\documentclass[a4paper,twocolumn,11pt,accepted=2025-06-10]{quantumarticle}
\pdfoutput=1
\usepackage[utf8]{inputenc}
\usepackage[english]{babel}
\usepackage[T1]{fontenc}
\usepackage[numbers, compress, sort]{natbib}

\usepackage{bm}
\usepackage{amsmath}
\usepackage{amssymb}
\usepackage{mathtools}
\usepackage{hyperref}
\usepackage[llbracket,rrbracket]{stmaryrd}

\usepackage{graphicx}

\usepackage{dcolumn}
\usepackage{microtype}
\usepackage{tikz}
\usetikzlibrary{quantikz}

\begin{document}

\title{Logical Noise Bias in Magic State Injection}

\author{Nicholas Fazio}
\email{nicholas.fazio@sydney.edu.au}
\author{Robin Harper}
\author{Stephen D. Bartlett}
 \affiliation{Centre for Engineered Quantum Systems, School of Physics,\\ The University of Sydney, Sydney, New South Wales 2006, Australia.}

\maketitle

\begin{abstract}
Fault-tolerant architectures aim to reduce the noise of a quantum computation. Despite such architectures being well studied a detailed understanding of how noise is transformed in a fault-tolerant primitive such as magic state injection is currently lacking. We use numerical simulations of logical process tomography on a fault-tolerant gadget that implements a logical $T = Z(\pi/4)$ gate using magic state injection, to understand how noise characteristics at the physical level are transformed into noise characteristics at the logical level.
We show how, in this gadget, a significant phase ($Z$) bias can arise in the logical noise, even with unbiased noise at the physical level. While the magic state injection gadget intrinsically induces biased noise, with extant phase bias being further amplified at the logical level, we identify noisy error correction circuits as a key limiting factor in the circuits studied on the magnitude of this logical noise bias. Our approach provides a framework for assessing the detailed noise characteristics, as well as the overall performance, of fault-tolerant logical primitives.
\end{abstract}

\section{\label{sec:intro}Introduction}
Quantum computation has the potential to solve a number of challenging problems, but current quantum devices are highly vulnerable to errors which limit their usefulness. Fault-tolerant architectures will allow for large-scale quantum computations even in the presence of noise~\cite{camp17, bever22}. While fault-tolerant quantum computing remains a long-term ambition, a number of recent experiments are exploring the foundations of fault-tolerant logic, demonstrating potential pathways for large-scale quantum computation~\cite{rist20, acha23, ande20, sund22, honey21, post22, marq22, bluv23}.

There are a variety of approaches to performing quantum logic fault-tolerantly. 
One prevailing approach is to construct families of error correcting codes with an error threshold~\cite{camp17}. 
If the physical error rate is below some threshold value, together with some other mild assumptions, then logical operations can be performed with arbitrarily small error rates through the use of quantum error correction. 
Threshold values have become a common metric for the performance of error correction schemes and fault-tolerant architectures, alongside other key considerations such as the locality of operations, connectivity of qubits, and spacetime-overheads. As a result of this emphasis on thresholds, research on quantum gate operations has focused on overall error rates, quantified by measures such as infidelity, diamond norm, or unitarity~\cite{evan22, blum17, beal18, huan19, iver20, wall15}, which can be readily compared against thresholds. 
Such a focus often ignores other characteristics of the physical noise and their effect on logical performance.

A recent and very fruitful line of investigation explores how to tailor quantum codes and architectures to features of the noise at the physical level, such as bias, coherence, or correlations across space and time. 
Such tailored codes have been shown to significantly impact performance~\cite{boni21, darm21, darm22}, beyond what is expected of generic noise models. 
While this research has improved our understanding of how noise characteristics impact fault-tolerance, we still have very limited understanding of the resultant noise at the logical level, and how this affects both performance and architecture design. 
An implicit assumption in prior works is that the characteristics of logical noise will reflect those of physical noise. For instance, if the physical noise is incoherent we would expect incoherent logical noise, and similarly that coherent physical noise would lead to coherent logical noise. 
However, it has been shown that measuring error syndromes decoheres physical errors~\cite{beal18}, reducing the coherence of noise at the logical level.
It is still not well understood how more general noise characteristics are relevant in the context of error correction~\cite{iyer22, jain23}.
Recent work in quantum sensing has shown that using error correction can have a notable impact on the character of noise~\cite{rojk22}. 
Let us consider, in particular, the performance of logical gates that are implemented using magic state injection---a primitive used in many fault-tolerant architectures to achieve universal quantum computation~\cite{gott99, cody13phd,bever20,bever21, brav04}. 
The prolific usage of this gadget in quantum logic issues an imperative for understanding the logical noise associated with its use. This is especially important in this emerging period where experiments are beginning to investigate fault-tolerance and the implementation of schemes for magic state injection~\cite{honey21, post22, marq22}.

In this work, we investigate how the characteristics of noise at the physical level are manifest in noise characteristics at the logical level when performing logical gates using magic state injection. Particularly, we focus on biased Pauli noise, and a fault-tolerant construction for performing logical $T$-gates (a non-Clifford gate) based on the Steane code. 
We find that this implementation of the logical $T$-gate leads to an inherent noise bias at the logical level, even for unbiased (depolarising) noise at the physical level. 
A bias at the physical level towards dephasing ($Z$ noise) is significantly amplified at the logical level. We undertake our investigation using numerical simulations of logical process tomography on this fault-tolerant gadget. These simulations allow us to isolate and identify the components of the logical $T$-gate gadget that lead to biased logical noise and those that limit the bias.
Although we use the Steane code to drive our investigation, we focus on the aspects of logical noise characteristics that apply more broadly. 
The factors that contribute to the biasing of logical noise here are also relevant in larger codes implementing magic state injection, where simulations are less tractable.
We consider the implications of our results for implementations of state injection, suggesting how logical bias can be managed, which will be especially relevant for magic state distillation schemes~\cite{liti19, brav04, mek13, brav12, cody13, camp17synth, camp18, haah17, haah18}. 

We present our results as follows. In Sec.~\ref{sec:methods}, we describe the noise characteristics and noise models that we focus on in our simulations. In Sec.~\ref{sec:results} we present the results of our simulations, exploring how physical noise characteristics impact logical noise, and in particular how noise bias affects each individual component of state injection. Finally, in Sec.~\ref{sec:discussion} we discuss how these features impact the use of state injection in schemes of fault-tolerance, and suggest approaches to manage and leverage the bias of noise.

\section{Methods} \label{sec:methods}
We use numerical simulations to predict the characteristics of logical noise associated with fault-tolerant primitives. The details of these numerical simulations are outlined throughout this section, as are our assumptions. 

We start by reviewing notation and terminology in Sec.~\ref{sec:prelim}. In Sec.~\ref{sec:noise} we define noise bias and outline how our noise models are specified and simulated. Finally, in Sec.~\ref{sec:f-t} we review the schemes that we use to perform logical tomography of encoded magic state injection.

\subsection{Gates and Noise Metrics}\label{sec:prelim}
Here we review standard terms and metrics used throughout this work.

We use the word \textit{fault} to refer to a state where an operation has been implemented with an error and use the word \textit{faulty} to describe an operation or collection of operations that are subject to faults.
An \textit{error} is the specific operation that follows a faulty gate and the \textit{weight} of an error is the number of qubits it acts non-trivially upon.
Such an error can be propagated through a circuit, conjugating the error operation by the circuit components, and we also refer to the resultant propagated operation as an error.
This propagated error typically has a different weight to the unpropagated error.
The word \textit{error} may also be used to refer to a collection of errors in which case the weight refers to the sum of the weights of the product of errors at each distinct time step.
The \textit{order} of an error is the number of independent gate faults that gave rise to the nominated error.

Phase operators of the form $Z(\theta) = e^{-i\frac{\theta}{2} Z}$ will be a focus of this work, especially the non-Clifford operator 
$$T = Z(\pi/4) = e^{-i\frac{\pi}{8} Z} = e^{-i\pi/8}\begin{bmatrix} 1 & 0\\ 0& e^{i\pi/4}\end{bmatrix}$$
commonly known as the \textit{T-gate}~\cite{camp17}, whose logical process we implement through our simulated circuits.
We will also refer to the \textit{S-gate}, $S = Z(\pi/2) = T^2$, a Clifford operator that we use as an adaptive correction.
Associated with the $T$-gate is the magic state $\ket{T} = T \ket{+} \sim \ket{0}+e^{i\pi/4} \ket{1}$ which is stabilised by the Clifford operator $TXT^\dag = SX = (X+Y)/\sqrt{2}$. 

We represent quantum operations as completely-positive trace-preserving maps and define the \textit{noise} of a faulty operation $\tilde{\mathcal{U}}$ to be the map $\mathcal{E}$ such that $\tilde{\mathcal{U}} = \mathcal{E}\circ \mathcal{U}$, where $\mathcal{U}$ is the ideal operation. 
The \textit{infidelity} of an operation refers to the average gate infidelity of the noise channel to the identity~\cite{niel11, beal18},
\begin{equation}
r(\mathcal{E}) = 1-F(\mathcal{E}) = 1-\int d\psi \langle\psi | \mathcal{E}(|\psi\rangle\langle\psi|)|\psi\rangle.
\end{equation}
The infidelity of a process can be determined experimentally using process tomography~\cite{niel11, GST, evan22}. 
Operations can be physical, acting on physical qubits, or logical, acting on logical qubits. 
When referring to \textit{logical noise}, we are always referring to process noise that acts entirely within a logical subspace.

\subsection{Noise Models}\label{sec:noise}
The exact noise characteristics that we explore are essential details for this work. We define those characteristics and how we perform our simulations of noisy circuits in this section. Through our simulated noise models we can understand how certain characteristics of physical noise influence the characteristics of logical noise in fault-tolerant gadgets. Magic state injection is one such gadget, and is an example where the physical-to-logical noise relationship is non-trivial. 

In order to simplify the description of general noise we use Pauli noise models. Restricting to this class of noise models also makes our simulations more tractable.
Work investigating how coherent noise impacts fault-tolerant quantum computation~\cite{beal18, huan19,iver20} has shown that the proportion of noise that is coherent is expected to decrease with increased distance, making Pauli noise models relevant for investigating large-scale fault-tolerant quantum computation.
Moreover, Pauli frame randomisation techniques such as randomised compiling can be used to transform general noise models into stochastic Pauli noise~\cite{wall16, erha19, harp19}.

For the $n$-qubit Pauli group $\mathcal{P}_n$, we can express any stochastic Pauli noise model acting on $n$ qubits in the form
\begin{equation}
\mathcal{E}(\rho) = \sum_{P_i\in \mathcal{P}_n} p_i P_i\rho P_i^\dag
\end{equation}
where $p_i$ is the classical probability of Pauli $P_i$ occurring. Any such noise model can be fully described with at most $4^n-1$ parameters.

The real noise of a device involves many parameters to fully classify, even when performing randomised compiling. However, many of these parameters are not relevant in practice, with a much smaller number of parameters accounting for a large portion of the noise. A relevant example of such a parameter is the characteristic of noise called the \textit{bias}~\cite{tuck18, tuck20, puri20} and it is the focus of this work.

\subsubsection{Pauli Noise Bias} \label{sec:bias}

Since we are employing Pauli noise models, measures such as the unitarity~\cite{wall15, iver20} or coherence angle~\cite{iver20}, which concern coherent noise, are not relevant here.
The bias of noise~\cite{tuck18, tuck20, puri20}, however, is a relevant noise characteristic to consider.
For instance, it has been shown to be a characteristic of noise that can increase threshold error rates~\cite{darm21, boni21, tuck18, tuck20}, therefore, having relevance for fault-tolerance overheads.

We define a general notion of Pauli noise bias, which remains consistent with previous definitions~\cite{tuck18, tuck20, puri20}, as follows.
Given the set of Pauli errors acting on $n$ qubits $\mathcal{P}_n\backslash\{\mathbb{I}\}$, we can partition the errors into two sets. The first set $\mathcal{Q} \subset \mathcal{P}_n\backslash\{\mathbb{I}\}$ we call the set of high-rate errors. The second set is its complement $\mathcal{Q}^C\subset \mathcal{P}_n\backslash\{\mathbb{I}\}$, which we call the set of low-rate errors. From this partitioning we define the $\mathcal{Q}$-bias of noise
\begin{equation}
\eta_n^\mathcal{Q} = \frac{\sum_{Q \in \mathcal{Q}} p_Q}{\sum_{Q' \in \mathcal{Q^C}} p_{Q'}}
\end{equation}
where $p_Q$ is the probability of a specific Pauli error $Q$ occurring. The total error rate $p$ is the sum of error probabilities over all Pauli errors $P\in \mathcal{P}_n\backslash\{\mathbb{I}\}$. For stochastic noise, the bias takes values $\eta_n^\mathcal{Q}\in [0,\infty)$ with $\eta_n^\mathcal{Q} \to\infty$ indicating noise completely biased to $\mathcal{Q}$ and $\eta_n^\mathcal{Q}=0$ indicating that errors from $\mathcal{Q}$ do not occur. We use the term \textit{extent} of bias to refer to the numerical value of~$\eta_n^\mathcal{Q}$.

Since we are considering quantum circuits that include two-qubit entangling gates, we would also like a measure of noise bias that specifically applies to correlated noise over two qubits. Throughout this manuscript we will use \textit{correlated noise} to refer to the kind of two-qubit noise that an entangling gate can introduce to both qubits it operates on. The measure of bias for correlated noise should also be consistent with existing parametrisations of single-qubit bias~\cite{tuck18,tuck20}, such as $Z$-bias
\begin{equation}
\eta_1^Z = \frac{p_Z}{p_X+p_Y}
\label{eq:singlequbit}
\end{equation}
One can define similar quantities for $X$- and $Y$-bias. We analogously define single-qubit $Z$-bias at the logical level, denoted by $\eta_1^{Z_L}$.

For two-qubit noise bias, we define the $P$-bias $\eta^P_2$ to be the ratio of error rates in which the numerator has all errors that are tensor products of $P$ and $I$ exclusively, and all other Pauli errors are in the denominator. In the case of two-qubit $Z$-bias,
\begin{equation}
\eta^Z_2 \vcentcolon= \eta^{\{Z_1,Z_2,Z_1Z_2\}}_2 = \frac{p_{Z_1}+p_{Z_2}+p_{Z_1Z_2}}{p-p_{Z_1}-p_{Z_2}-p_{Z_1Z_2}}
\end{equation}
Defining two-qubit bias in this way guarantees that the concept of high $Z$-bias still implies $Z$ errors occur at high-rate, but also that $X$ and $Y$ errors never occur at high-rate. As an example, $Z_1Y_2$ would occur at low-rate.

For depolarising noise all Pauli errors have equal probability and so for single-qubit depolarising noise we have the following values for noise bias: $\eta_1^X =\eta_1^Y=\eta_1^Z= 1/2$. For two-qubit depolarising noise the values for noise bias are different, namely $\eta_2^X =\eta_2^Y=\eta_2^Z= 3/12$. 

Further details on bias can be found in App.~\ref{sec:biasdef}.

\subsubsection{Simulated Noise Models}\label{sec:sim}
We need noise models that are consistent and comparable with each other in order to investigate how the characteristics of noise affect fault-tolerant gadgets. We specify our Pauli noise models in this subsection.

We vary our Pauli noise channels by the rate of error and the extent of noise bias. For error rate~$p$ and bias parameter $\eta_n^\mathcal{Q}$ we define
\begin{equation}
\begin{split}
\mathcal{N}(\eta^\mathcal{Q}_n, p)(\rho) & = (1-p)\rho + \frac{p\eta_n^\mathcal{Q}}{(\eta_n^\mathcal{Q}+1)} \sum_{Q\in\mathcal{Q}}\frac{1}{|\mathcal{Q}|}Q\rho Q\\ 
& +\frac{p}{(\eta_n^\mathcal{Q}+1)} \sum_{Q'\in\mathcal{Q}^C}\frac{1}{|\mathcal{Q}^C|}Q'\rho Q'
\end{split}
\end{equation}
to be the $n$-qubit noise channel such that the total error rate is equal to $p$, the $\mathcal{Q}$-bias of noise is equal to the parameter $\eta_n^\mathcal{Q}$, and the individual Pauli error probabilities are such that all high-rate errors have equal probability and all low-rate errors have equal probability. 

In our simulations we implement the controlled-$X$ gate, also known as the CNOT gate, as an operator sustaining biased noise. The other operations in our simulated circuits are performed ideally. Generally each operation on a physical device will have its own noise characteristics. However, entangling gates are typically among the noisiest operations on a device~\cite{heus23, acha23}, especially compared to single-qubit gates, and in the circuits we consider they are the most common gates. We expect that the assumption made here should be broadly applicable, addressing the predominant factors that contribute to logical noise characteristics. Even so, we will later discuss how our results could change if we relax this assumption. Since we are using Pauli noise models, Pauli errors from other sources such as single-qubit gates are equivalent to CNOT Pauli errors with the exact same scaling in the error rate $p$. So we can infer what impact the neglected terms would have through our assortment of noise models.

Our noisy CNOTs independently experience a biased Pauli noise channel over two-qubits. We define our noisy CNOT as
\begin{equation}
\tilde{C}_X(\rho) = \mathcal{N}(\eta^\mathcal{Q}_2, p)(C_X(\rho))\ ,
\end{equation}
where the control qubit of the CNOT operator $C_X$ corresponds to the first qubit in $\mathcal{Q}$. For example, if $\mathcal{Q}=\{Z_1,Y_2,Z_1Y_2\}$, then $Z$ occurs with high-rate on the control qubit of the CNOT and $Y$ occurs with high-rate on the target qubit. Accordingly, $p_{Z_1}$, $p_{Y_2}$ and $p_{Z_1Y_2}$ are the highest probabilities among Pauli errors.

We perform our simulation by placing the initial Pauli errors in the circuit, conjugating these errors to the end of the circuit, and then computing the impact of the error on the ideal logical state before measurement.
For each noise model, we simulate all error terms where the coefficient has a factor of up to $p^3$ or fewer, that is, with order less than or equal to 3. Any terms with order 4 or higher are truncated.
Preliminary simulations of higher order terms suggested that their contributions were negligible due to the low distance of the code and relatively small number of CNOTs used. The qualitative analysis of this work is unchanged by this approximation.

Note that $p$ is the rate of individual CNOT \textit{faults}, so correlated two-qubit errors occur at a rate proportional to~$p$. Hence, the weight of an error can be twice the number of two-qubit gate faults, that is twice the order. 
For circuits in low-distance codes, such as those used for state preparation and error correction, these correlated errors can be challenging to deal with due to their high weight, relative to the code distance.
However, it is important that we consider correlated errors because they commonly occur in the entangling gates of physical devices, and may even be the dominant error.

Through our biased noise models we can investigate the predominant sources of logical noise in a tractable way. We can vary both the rate of error and the bias of physical noise to discern how changes in the noise characteristics impact logical expectation values.

\subsection{Fault Tolerance}\label{sec:f-t}

Quantum devices are prone to errors so any form of computation that makes use of them must also tolerate the ongoing faults of the device. Key to fault tolerance is that the processes of detecting and correcting errors can be used even if these processes also introduce noise.
If all errors below some required order (the number of independent faults that cause the error) do not culminate in a logical error, regardless of which operations are faulty, we can call such a scheme \textit{fault-tolerant}.
Recent research has shown that the nature of noise on a device can have a considerable impact on thresholds~\cite{boni21, darm21, darm22}, but the impact of noise characteristics on logical noise~\cite{beal18, huan19, iver20} is largely unexplored.

\begin{figure}[t]
\centering
\includegraphics[width=0.22\textwidth]{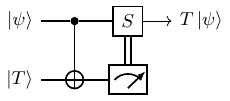}
    \caption{A non-Clifford $T$-gate can be performed with a resource magic state $\ket{T}$ using magic state injection. The injection process is adaptive, applying an $S=Z(\pi/2)$ correction when $\ket{T}$ is measured in the $\ket{1}$ state, and no correction otherwise.}
    \label{fig:msi}
\end{figure}

We choose to use the Steane code~\cite{camp17, steane96, goto16, reich18no} to investigate logical noise characteristics due to its simplicity and widespread use in fault-tolerant constructions. The Steane code possesses a number of logical operations that can be implemented transversally, which is the simplest form of a fault-tolerant gate. However these transversal operations are all elements of the Clifford group, and so to achieve a universal gateset~\cite{east09} the transversal logic gates must be supplemented with gates that are implemented using an alternative fault-tolerant construction.

Magic state injection~\cite{brav04}, outlined in Fig.~\ref{fig:msi}, is an important primitive of quantum logic that can be used to implement the non-Clifford $T$-gate. In conjunction with the Clifford group the $T$-gate forms a universal gateset, so we can use magic state injection to compensate for the non-transversal gates of the Steane code. High-fidelity magic states can be used to implement high-fidelity $T$-gates, making encoded magic state injection an important primitive for fault-tolerance~\cite{gott99}. 

Performance metrics already exist for magic states~\cite{vida99, hein19, sedd19, bever20}, but to our knowledge there has been no work that investigates how the characteristics of noise, such as bias, affect the logical noise on a $T$-gate performed with an injection gadget.
We explore this particular logical gadget for the Steane code using logical process tomography, a scheme that can be used to fault-tolerantly characterise logical noise in experiment. Using fault-tolerant state preparation schemes for the Steane code~\cite{goto16, cham19}, and fault-tolerant logical measurements, we have the means to perform logical process tomography for a logical $T$-gate, which we simulate numerically to probe logical noise characteristics.

\subsubsection{Steane Code}

\begin{figure}[t]
\centering
\includegraphics[width=0.27\textwidth]{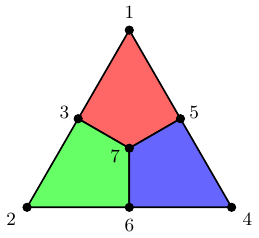}
\caption{The Steane code is a seven qubit stabiliser code. Each numbered vertex represents a qubit of the code. Corresponding to each coloured face there are $X$ and $Z$ stabiliser generators that act on all incident vertices.}
\label{fig:steane}
\end{figure}

Our simulations are performed in a $\llbracket 7, 1, 3\rrbracket$ code called the Steane code~\cite{camp17, steane96, goto16,reich18no}, depicted in Fig.~\ref{fig:steane}. It encodes one logical qubit in seven physical qubits. The transversal $X$, $Y$ and $Z$ logical operators all have a minimal weight of three, meaning that the code can detect up to two physical errors and can correct up to one~\cite{gott97}.
For fault-tolerance in the Steane code we will require that any single fault should not culminate in a logical error.

The entire logical Clifford group can be performed transversally on the Steane code. Therefore, together with a means to perform fault-tolerant logical $T$-gates using magic state injection, a universal set of fault-tolerant gates can be implemented for this code.

The aforementioned properties make the Steane code a viable candidate for the demonstration of a fault-tolerant gateset. Since it is small and implementable in a planar layout using nearest-neighbour entangling gates, it is accessible for near-term experiments~\cite{honey21, post22, bluv23}. While codes with a larger distance are likely to be needed for practical fault-tolerant approaches, we can use the Steane code to establish the current performance of encoded operations and how characteristics of physical noise impact logical noise in current devices.

\subsubsection{Logical Tomography}

Logical process tomography~\cite{post22, ridd23} is the technique that we use to identify logical noise characteristics in this work. Essentially, by using logical state preparation and measurement we can perform tomography on a logical process analogously to how process tomography is performed on a physical process. We also require that all the components of logical tomography are fault-tolerant, as defined earlier in Sec.~\ref{sec:f-t}.

\begin{figure}[b]
\centering
\includegraphics[width=0.42\textwidth]{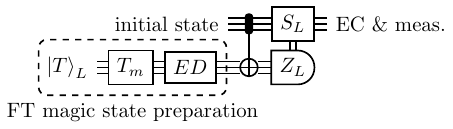}
    \caption{Logical process tomography of a $T$-gate implemented through magic state injection. Fault-tolerant magic state preparation is a fundamental part of the gadget. For the Steane code specifically, magic state preparation can be done via a state check $T_m$ and error detection $ED$. By preparing initial logical stabiliser states~\cite{goto16}, and by performing final error correction~\cite{reich18no, reich18two} with logical measurements, the logical process matrix of the $T$-gate can be reconstructed. We simulate such a circuit encoded in the Steane code (see App.~\ref{sec:tomography} for implementation details).}
    \label{fig:err-circ}
\end{figure}

We incorporate logical process tomography directly into the circuits that we simulate, encompassing a logical $T$-gate (implemented via magic state injection) with fault-tolerant state preparation and measurement. 
We perform our noisy tomography simulations in the Steane code, for which the abstract circuit structure is depicted in Fig.~\ref{fig:err-circ}.

We use the state preparation schemes of Goto~\cite{goto16} and Chamberland~\cite{cham19} to prepare logical states for the Steane code, preparing several different initial states for use in logical tomography. Further considerations on these schemes with respect to the noise models of this work can be found in App.~\ref{sec:state-prep}. In order to better tolerate correlated noise from faulty CNOTs we modify existing flagged stabiliser measurement schemes~\cite{yoder17, reich18two, reich18no} for use in state preparation, detailed separately in App.~\ref{sec:err-corr}. We also use the pre-existing schemes to measure stabilisers for error correction, which is also outlined in App.~\ref{sec:err-corr}.

The results of this work are obtained by simulating these noisy logical tomography circuits and using the logical data to derive the noise of the logical $T$-gate process. We consider logical state preparation, stabiliser measurements and logical measurements to all be essential components in our simulations of noise, motivated by how the noise of such circuits would impact the logical tomography results of experiment.
Unless otherwise stated, all CNOTs of the circuit are faulty including those used for tomography, that is, including those CNOTs involved in state preparation and stabiliser measurements.

The overall circuit structure, processing of measurement outcomes, and implementation of adaptive corrections, are details that are kept consistent across all simulations. The only difference between each simulation is the noise strength of the noise model experienced by the circuit. Full details on the treatment of measurement outcomes for logical tomography can be found in App.~\ref{sec:tomography}.

\section{Results}\label{sec:results}

Through a series of numerical simulations we will interrogate how noise characteristics at the physical level manifest in the characteristics of noise at the logical level for logical $T$-gates implemented via magic state injection using the Steane code. 
From this we discern the main mechanisms that influence the bias of the noise at the logical level. 
Our findings demonstrate that bias can be amplified at the logical level and identify the noise associated with error correction as playing a pivotal role in determining the characteristics of logical noise. 

We simulate logical tomography of logical $T$-gates implemented in the Steane code using state injection, utilising fault-tolerant schemes for state preparation and measurement~\cite{reich18no, reich18two, goto16, cham19}. We apply biased Pauli noise models to our simulated circuits in order to investigate how physical noise characteristics impact logical noise. We can isolate how different aspects of magic state injection influence the characteristics of logical noise by varying the parameters of our physical noise model, namely the physical rates of error, different kinds of physical noise bias, and which circuit components of the simulation are faulty. The full specifications of our simulations can be found in Sec.~\ref{sec:methods}, including how we define bias and parameterise our noise models.

First we consider the impact that physical noise bias has on logical process infidelity in Sec.~\ref{sec:inf}. From there we focus on the relationship between physical noise bias and logical noise bias. In Sec.~\ref{sec:biased}, we investigate why the logical $T$-gate implemented via state injection has intrinsically $Z$-biased noise, and highlight the noise-transforming properties that are inherent to magic state injection. Finally, in Sec.~\ref{sec:constraints} we examine how the type of physical noise bias matters in the state injection gadget, highlighting how noisy error correction limits the extent of logical noise bias in logical gates.

\subsection{Biased Physical Noise in the Steane Code} \label{sec:inf}

\begin{figure}[b!]
\centering
\includegraphics[width=0.5\textwidth]{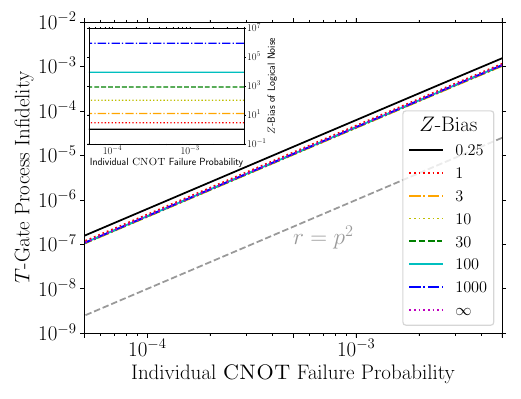}
\caption{Process infidelity of a logical $T$-gate implemented via magic state injection for a spectrum of $Z$-biased noisy CNOTs. The $Z$-bias of physical noise $\eta_2^{Z}$ spans from two-qubit depolarising noise at $\eta_2^{Z} = 0.25$ to pure physical $Z$ noise $\eta_2^Z\to\infty$. There is a quadratic reduction in logical process infidelity when compared to the physical error rate, which is the expected scaling when performing error correction in a distance-three code. Logical $Z$-bias $\eta_1^Z$ is largely unaffected by changes in the physical error rate.}
\label{fig:zinfid}
\end{figure}

The Steane code has distance $d=3$ and so can correct any single-qubit error. Since we use fault-tolerant circuits and implement logical gates transversally, we expect the logical infidelity to scale quadratically with the rate of physical error, which we see verified in Fig.~\ref{fig:zinfid} for all the Pauli noise models considered. Aside from the expected scaling, we note that the effect of physical noise bias on the logical infidelity is quite minor. We found this weak dependence on physical noise bias to hold in all the noise models considered within this text, beyond the noise models considered in this section.

The reason we do not see a strong relationship between the physical noise bias and the logical process infidelity is due to properties of the Steane code~\cite{robe17}, which is no better suited for biased noise than unbiased noise. 
To more directly exploit bias we could employ alternate codes tailored for biased noise, but we do not investigate that in this work.

However, we still observe a slight dependence of the logical infidelity on physical noise bias. We can understand this detail by considering the relative frequencies between different kinds of error. We qualitatively describe this as follows: 
As we change the physical bias, we also change the predominant ways that physical noise accumulates into logical noise.
At low bias, there are a large number of error terms, each with low probability, that contribute to the logical error. At high bias, a smaller number of error terms contribute, each with a higher probability. 
As we transition between these two regimes the logical infidelity changes as the relative contribution of each error term changes.
Regardless of bias, the sum probability of physical error remains the same. So unless particular error terms are far more likely to be successfully error corrected than others of the same weight the change to logical infidelity will be minimal, which is the case here as previously stated.

The inset of Fig.~\ref{fig:zinfid} indicates that the logical noise bias changes minimally with the rate of physical error. This is once again a consequence of the Steane code. 
Since the Steane code can correct weight-one errors, second order error terms are the lowest order contribution to logical noise.
Due to the small number of qubits in the Steane code and relatively low number of CNOTs used, these lowest order error terms also contribute to the vast majority of the logical noise.
In the context of Eq.~\ref{eq:singlequbit} specifically,  the largest contribution to the logical error rates in both the numerator and the denominator will be terms that take the form $cp^2$.
By factoring such terms to the front of the fraction we get a constant that is grossly indicative of the logical noise bias, with terms of third order and higher becoming less and less relevant with smaller rates of error, that is, $\eta_1^{Z_L} = p_{Z_L}/(p_{X_L}+p_{Y_L})\sim (ap^2+O(p^3))/(bp^2 + O(p^3))\sim \frac{a}{b} (1+O(p))$.
Nonetheless, for larger codes with larger distances this will not generally hold: lowest order error terms will be less numerous compared to the higher order terms, meaning that the coefficients of the higher order terms are larger and comparable to $1/p$, allowing for a more complicated dependence of logical noise bias on the physical error rate.

\subsection{Physical-to-Logical Bias} \label{sec:biased}
\begin{figure}[t]
\begin{center}
\includegraphics[width = 0.48\textwidth]{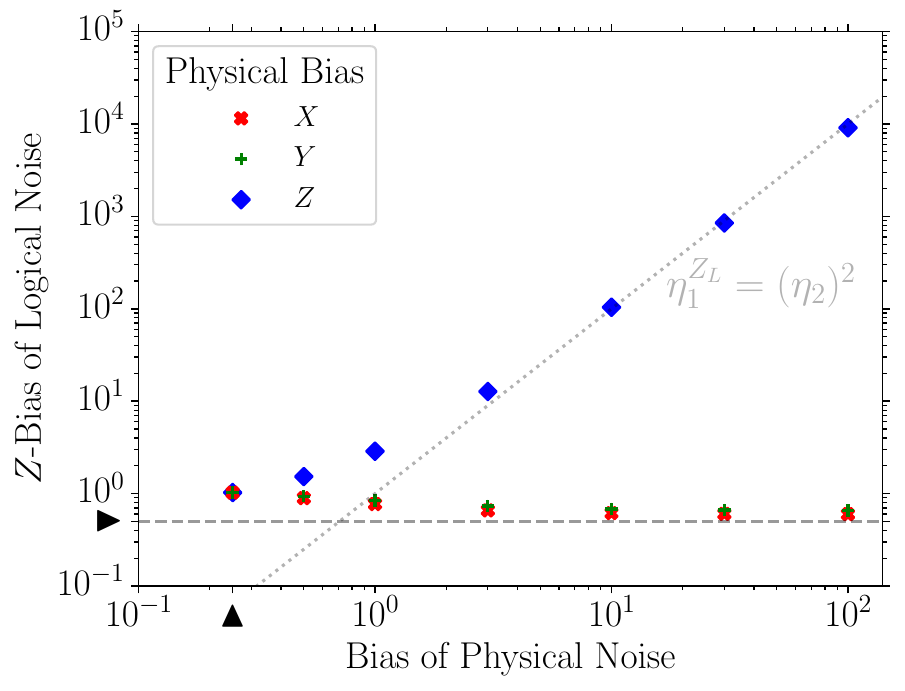}
\end{center}
\caption{Logical $Z$-bias while varying the bias of CNOT noise. Simulations are performed at error rate $p = 5\times 10^{-3}$. Marked by $\blacktriangle$ are the bias values for depolarising noise. The logical $Z$-bias is always greater than the bias for depolarising noise. The $Z$-bias of logical noise has an asymptotically quadratic relationship with the $Z$-bias of physical noise. Highly $X$- or $Y$-biased physical noise tends to logical depolarising noise.}
\label{fig:zbiased}
\end{figure}

We now turn to how physical noise bias impacts logical noise bias.
In Fig.~\ref{fig:zbiased}, we consider how the $Z$-bias of logical noise in our logical $T$-gate gadget is affected by different physical noise characteristics. 
Examining the case of $Z$-biased physical noise first, there is a clear quadratic relationship between the physical noise bias $\eta_2^Z$ and the bias of the logical process noise $\eta_1^Z$. This relationship is closely related to the code distance of the Steane code.

Explicating on this relationship, the most likely logical errors are those which come from high-rate physical errors. Since the Steane code's distance is three, the lowest weight error that can become a logical error is two. 
Since all first order errors are detected or corrected in our gadget, the most likely weight-two errors will occur when two independent sources of noise both introduce high-rate errors. 
For $Z$-biased physical noise, such high-rate error terms have a coefficient quadratic in $\eta_2^Z$ and are only composed of physical $Z$~errors, which shall become logical $Z$~errors. Hence, as the physical $Z$-bias increases, the most likely logical $Z$~errors scale quadratically in $\eta_2^Z$ while all other logical error terms have a decreasing likelihood, resulting in an amplification of bias at the logical level.

Strikingly, when the bias at the physical level is in the~$X$ or~$Y$ direction we still see a weak $Z$-bias at the logical level.
Since this $Z$-bias is rather small, we may effectively think of this as logical depolarising noise, which is rather different from seeing $X$- or $Y$-bias at the logical level.
This inherent $Z$-biasing occurs when the physical noise is depolarising, all the way through to highly $X$- or $Y$-biased noise, that is, when only $X$ and $Y$ errors are being introduced to the circuit.

\begin{figure}[b!]
\centering
\includegraphics[width=0.49\textwidth]{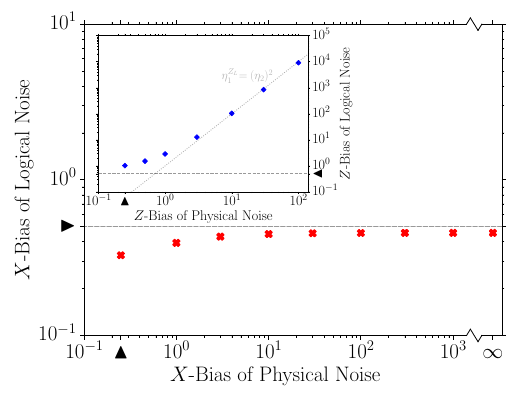}
\caption{Logical $X$-bias against physical $X$-bias for fixed rate of error $p = 5\times 10^{-3}$. Marked by $\blacktriangle$ are the bias values for depolarising noise. Although the logical $X$-bias marginally increases with physical $X$-bias, it plateaus below the value for depolarising noise. Even with pure $X$ physical noise, the logical noise is not $X$-biased. Contrast this with physical $Z$-bias versus logical $Z$-bias.}
\label{fig:comparebias}
\end{figure}

\begin{figure}[t]
\begin{flushleft}
(a)\\
  \raisebox{-0.5\height}{\includegraphics[width=0.2205\textwidth]{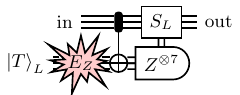}}
  $\to$
  \raisebox{-0.45\height}{\includegraphics[width=0.2156\textwidth]{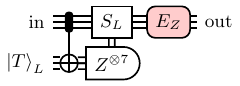}}\\
(b)\\
  \raisebox{-0.5\height}{\includegraphics[width=0.2205\textwidth]{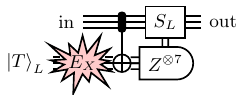}}
  $\to$
  \raisebox{-0.45\height}{\includegraphics[width=0.1813\textwidth]{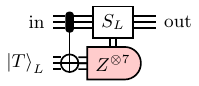}}
\end{flushleft}
\caption{Pauli errors introduced by magic state preparation can only amount to phase errors on the computational qubits. In (a) $Z$~errors on the magic state spread to $Z$~errors on the computational qubits with the same weight and syndrome. Even though $X$ errors do not spread to the computational qubits in (b), they affect the logical measurement outcome. Depending on the weight and syndrome of the $X$ error the logical measurement can fail and cause a misapplication of the phase gate $S = Z(\pi/2)$, introducing a logical phase error.}
\label{fig:z-origin}
\end{figure}

The same reasoning applied for $Z$-bias would suggest that we should see a similar amplification in logical bias with $X$- or $Y$-biased physical noise. However, as explicitly demonstrated in Fig.~\ref{fig:comparebias}, this does not occur. Rather than scale quadratically, we find the bias of logical noise plateaus well before the logical noise becomes $X$-biased. Even when the physical noise is purely composed of $X$~errors, there is seemingly some additional source of $Z$~noise that causes the logical noise to be mixed. 

The quadratic scaling only occurs with~$Z$ due to mechanisms within magic state injection that convert arbitrary errors into $Z$~phase errors. So although the physical noise becomes increasingly $X$-biased, the second order $X$~error terms must also compete with other sources of logical $Z$~error. In contrast, the $Z$-bias of logical noise is unimpeded and can scale quadratically, for the reasons previously detailed.

We depict the cause of this mechanism in Fig.~\ref{fig:z-origin}.
Errors incurred during magic state preparation either spread as $Z$~errors to the computational qubits, or cause misapplications of the logical correction $S=Z(\pi/2)$.
Therefore, any type of Pauli noise on the logical magic state transfers to the computational qubits as phase noise.
If only magic state preparation were noisy, the logical process noise of the $T$-gate would be completely $Z$-biased regardless of the underlying physical noise model.
In such a case logical $X$~and $Y$~errors would still be present on the logical magic state itself, but these errors would manifest as phase errors on the computational qubits. 

Misapplied logical $S$ corrections not only result in logical phase noise but at the process level induce logical $Z$ noise across multiple runs.
Half the time a misapplication results in a logical $S^\dag T$, and the other half it results in a logical $ST$.
Therefore, since $S\rho S^\dag + S^\dag \rho S = \rho + Z\rho Z^\dag$, such equal probability misapplications result in a density operator with logical $Z$~errors, which occur at half the misapplication rate.

\subsection{Structure of the Gadget Noise} \label{sec:constraints}

\begin{figure*}[t]
\centering
\begin{flushleft}
\large (a)\hspace{0.5\textwidth}(b)\\
\end{flushleft}
\vspace*{-5mm}
\includegraphics[width=0.46\textwidth]{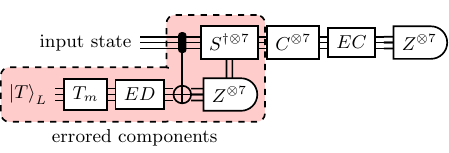}
        \hfill    
\includegraphics[width=0.46\textwidth]{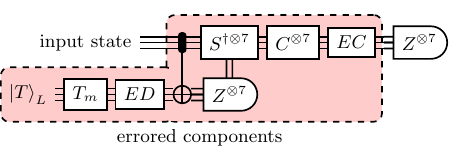}\\
    \includegraphics[width = 0.475\textwidth]{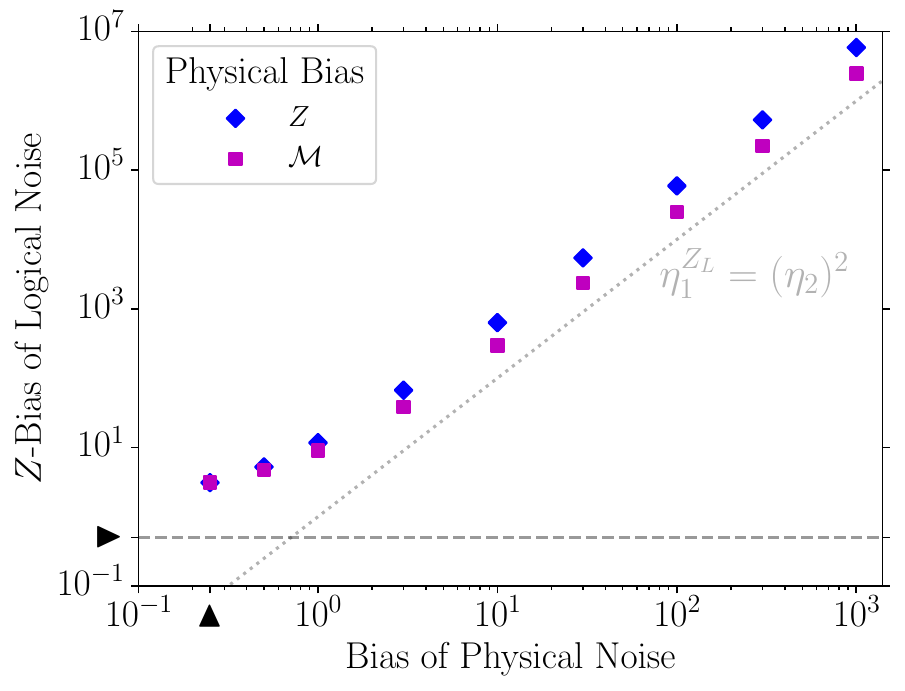}
        \hfill
    \includegraphics[width = 0.475\textwidth]{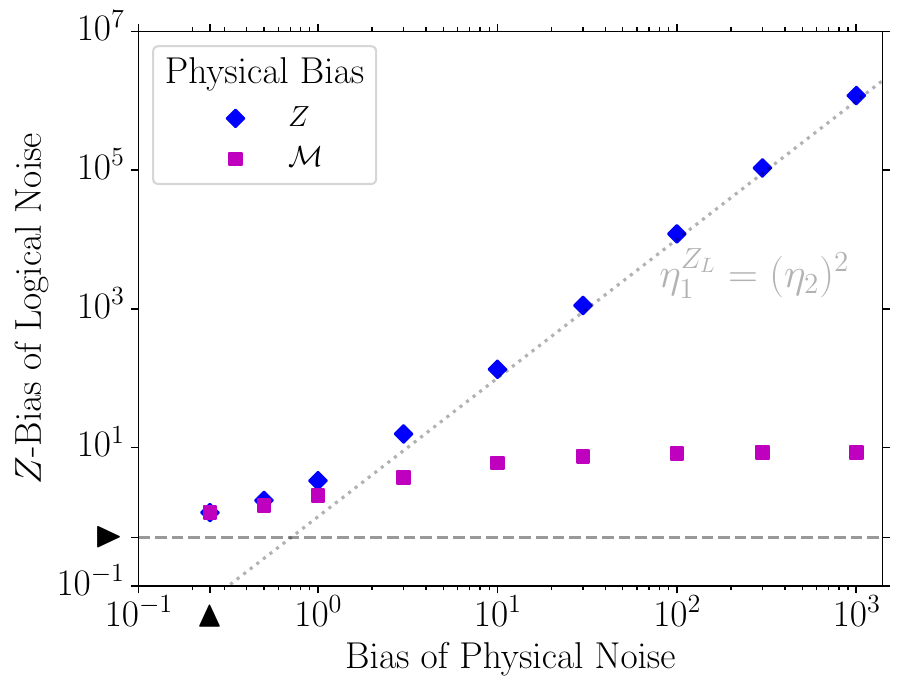}
\caption{Simulations of noisy logical tomography circuits encoded in the Steane code. Transversal Clifford gates $C^{\otimes 7}$ are used to rotate into the final logical basis for measurement. $Z$-biased physical noise $\eta_2^Z$ and mixed-biased physical noise $\eta_2^\mathcal{M}$ are plotted against the $Z$-bias of logical noise $\eta_1^{Z_L}$ for fixed rate of error $p = 5\times 10^{-3}$. Marked by $\blacktriangle$ are the bias values for depolarising noise. In~(a) stabiliser state preparation and error correction are ideal while the rest of the state injection is noisy. Despite the noise bias being mixed $\mathcal{M} = \{Z_1,X_2, Z_1X_2\}$, the $Z$-bias of logical noise scales quadratically with the bias of physical noise. In~(b) error correction is also included in the faulty components, causing the logical $Z$-bias to plateau as the $\mathcal{M}$-bias of physical noise increases. Although mixed physical noise introduces high-rate $X$ and $Z$ noise, it is faulty error correction that allows for the high-rate $X$ errors to limit the extent of logical $Z$-bias.}
\label{fig:z-control}
\end{figure*}

In previous sections we have seen how the structure of the magic state injection gadget transforms the characteristics of noise. These noise-transforming properties lead to an inherent $Z$-bias in the logical process noise, which is amplified by error correction if the physical noise is also $Z$-biased. However, the simple Pauli noise models of previous sections do not allow us to anticipate the logical noise bias from the physical noise bias, or how each component of the injection gadget contributes to the logical noise. We explore these questions in this section by investigating the contributing factors to the logical noise incrementally. By breaking down the observed logical noise into the contributions from individual components of the gadget we can understand how bias affects each component separately and how this culminates in the noise characteristics of the logical process. 

The results of Sec.~\ref{sec:biased} provide a starting point for this analysis. They demonstrate that noise of any character originating from magic state preparation leads to logical phase noise. That is, if magic state preparation was the only source of noise, the logical process would have noise that is completely $Z$-biased.

Looking at noise with a more complex structure will allow us to better understand how the type of bias impacts different components of the logical gadget and the logical noise. We will use a mixed-bias noise model where the high-rate set of Pauli errors is $\mathcal{M} = \{Z_1,X_2, Z_1X_2\}$, that is, as the extent of bias $\eta_2^\mathcal{M}$ increases, the noise introduces both $Z$ and~$X$ errors at high-rate. Notably, the control qubit of the CNOT gets biased towards $Z$ errors while the target qubit gets biased towards $X$ errors. $\mathcal{M}$-biased noise on a CNOT is equivalent to $Z$-biased noise on a Controlled-$Z$ gate,
\begin{equation}
\vcenter{\hbox{\includegraphics[width=0.19\textwidth]{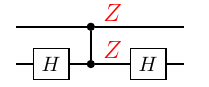}}} =\vcenter{\hbox{\includegraphics[width=0.1\textwidth]{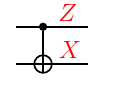}}}
\end{equation}
a type of bias that is motivated by experiment~\cite{mckay19, cong22, saha23}.

In Fig.~\ref{fig:z-control}(a), we consider the logical performance of our process when all the elements of magic state preparation and injection are noisy, but with stabiliser state preparation and error correction performed ideally so that we can focus on just the noise from magic state injection.
Previously we have considered noise in circuit components that would influence the logical outcomes in logical process tomography, but here we want to focus on logical process noise unburdened by logical state preparation and measurement noise.
In this setting, we find that for increasing physical $Z$-bias the logical $Z$-bias scales quadratically as we would expect, however, we also find the same quadratic scaling while increasing the extent of mixed-bias~$\eta_2^\mathcal{M}$.
Even though mixed-biased physical noise introduces both $X$ and~$Z$ errors at high rate, the logical noise only becomes biased to~$Z$. 
We would naively expect that the logical noise should be mixed like the physical noise. However, a distinction between logical $X$ and $Z$ arises due to the distribution of physical errors: high-rate $X$~errors are only being introduced to the logical magic state while all the $X$ and $Y$ errors that are introduced to the computational qubits are low-rate errors. 
As the mixed-bias of physical noise increases, the low-rate $X$ and $Y$ errors become even less likely, introducing a higher proportion of physical $Z$ noise onto the computational qubits. Simultaneously, the high-rate $X$ errors are converted into phase errors via the same mechanism as in Fig.~\ref{fig:z-origin}. Hence, mixed physical noise does not necessarily imply mixed logical noise, as seen in Fig.~\ref{fig:z-control}(a), the logical noise can be highly $Z$-biased.

However, when we introduce faulty error correction, the quadratic scaling of logical $Z$-bias is not maintained at large bias. From Fig.~\ref{fig:z-control}(b) we can see that the logical $Z$-bias plateaus with increasing mixed-bias. The important difference is that faulty error correction introduces a new source of high-rate $X$ noise. The mixed character of bias becomes relevant for our logical $T$-gate through the CNOTs involved in error correction. In the absence of faulty error correction, $T$-gates implemented with magic state injection are capable of having highly $Z$-biased noise, even under mixed noise models, but in general the characteristics of noise in error correction constrain the $Z$-bias of the logical process.
Despite this constraint, we can still see that the $Z$-bias remains fairly high in comparison to the noise models of previous sections where the logical $Z$-bias also plateaued, as in Fig.~\ref{fig:zbiased}. Although this previous figure includes noise in state preparation, by comparison it is clear how the high rate $Z$ errors that are introduced by $M$-biased noise are still prominent in the logical $Z$-bias.
So, the strong $Z$-bias that we would have seen if error correction were ideal still makes an important contribution to the overall logical bias, albeit tempered by error correction.

Ultimately, this demonstrates that the noise characteristics of error correction play a determinative role in the logical noise characteristics of magic state injection.
Since error correction is a fundamental component in fault-tolerance, all logical processes, including transversal gates, must contend with the noise of error correction and how it affects extant noise. 

\section{Discussion} \label{sec:discussion}

In this work we have set out to understand how the characteristics of physical noise affect logical noise in fault-tolerant primitives. 
We explore this question by simulating biased noise models in an encoded magic state injection circuit, allowing us to discern the predominant factors that influence the relationship between noise characteristics at the physical level and at the logical level.

By far the most salient feature of our simulations is the $Z$-bias of logical noise. 
This logical $Z$-bias is greatest when extant physical $Z$-bias is amplified by error correction, 
however, across all the noise models considered some level of logical $Z$-bias can be found due to the noise-transforming properties of the injection gadget. 
Bias-amplification is ultimately also a kind of noise-transformation, induced by the stabiliser code, but the injection gadget separately ensures that even in the presence of $X$-biased physical noise the logical noise is marginally $Z$-biased. 
In the analysis presented here, the logical noise specifically tends to~$Z$ because we chose to implement $T = Z(\pi/4)$; had we chosen another basis for our operator such as $X(\pi/4)$ or $Y(\pi/4)$, the bias introduced by the injection gadget would reflect that choice.
Likewise, conjugating the $T$-gate by a Clifford operator would realise an analogous situation.

These noise-transforming properties are an intrinsic feature of the injection gadget, which do not occur in transversal Clifford gates alone. 
While Clifford gates can change the type of bias, the overall extent of noise bias remains unchanged up to a Clifford frame.
In contrast, the extent of noise bias fundamentally changes in magic state injection because the errors on the logical magic state get transformed.
Consequently, since the gadget's noise-transformation only affects noise on the magic state, the structure of physical noise bias impacts the logical noise characteristics of magic state injection.
An example from this work occurs in simulations of mixed-bias noise, where high-rate $X$ errors predominantly arose on the logical magic state, but these high-rate $X$ errors did not prevent a high logical $Z$-bias from appearing.
This noise-transformation presents us with an opportunity; we can exploit the inherent noise-transforming properties to bring about highly biased logical noise.

Throughout this work there is a presupposition that high bias is worth pursuing. The reason for this is straightforward: a higher bias is preferable whenever the overall error rate is unchanged because we can tailor fault-tolerant schemes to take the bias into account. 
However this reasoning is usually applied in the context of physical noise bias rather than the logical noise bias investigated here. One way to take advantage of logical bias is through concatenated codes, with the bias informing which codes will be appropriate for concatenation. 
Since we are specifically considering $T$-gates within a logical code, the most natural context for this work is magic state distillation.

Magic state injection is a common gadget in magic state distillation, used to pass magic states onto the next round of distillation~\cite{camp17,liti19,mek13,brav12}.
Analysis of magic state distillation schemes commonly assumes that magic states are the faulty components in each distillation round, which corresponds to purely $Z$-biased process noise in our work. 
However, under certain schemes of magic state distillation~\cite{liti19} the presence of logical $X$~and $Y$~errors can have a significant impact on the performance of distillation. 
Since logical $X$~and $Y$~errors that are introduced by a $T$-gate (as opposed to on a magic state) do not commute with the non-Clifford phase operators of distillation, such errors can cause multiple phase operators to be misapplied simultaneously.
In other words, these errors are equivalent to correlated errors over multiple magic states, thus having a far greater impact on the performance of distillation than $Z$~errors. 
Litinski accounts for such errors by modifying the $X$-distance of code-patches~\cite{liti19}. 
Hence, the extent of $Z$-bias in noisy $T$-gates is of consequence for fault-tolerance overheads, making it an important metric in the performance of distillation schemes.

Naturally, the noise transformed by the injection gadget will also be affected by noise from later circuits. 
Whilst this work has shown that noise-transformation is indeed very relevant at the logical level, noisy error correction circuits will typically limit the extent of logical $Z$-bias emerging from the gadget by introducing unbiased noise.
Other fault-tolerant operations, such as transversal operators, are also influenced by noisy error correction circuits in the same way, so that without the noise-transforming properties of magic state injection the logical noise bias of these operators will be even more strongly determined by the noise from error correction.

Other sources of Pauli noise, such as noisy single-qubit gates, would have a similar impact on noise characteristics in magic state injection. These errors may likewise diminish the bias of logical noise or be converted into logical phase noise, albeit independently of the physical noise bias of CNOTs.
For the most part, state preparation and measurement (SPAM) errors would affect logical tomography rather than the logical noise of the $T$-gate.
Idling errors and other sources of intermediate noise are far more relevant to the characteristics of logical noise that we have investigated here since they will contribute noise systematically across the entire gadget, which is quite likely to affect the extent of logical bias.
However, as initially prefaced, if CNOTs are the noisiest operators on a device then noisy error correction will constitute the majority of the aforementioned factors for logical bias.

Although noise-transformation has the capability to bring about high logical bias, the constraint placed on bias by noisy error correction circuits suggests that the greatest challenge in the pursuit of high bias will be the way that error correction is implemented. 
This problem motivates the development of circuits for error correction that are bias-sensitive, preferring to introduce $Z$~errors in place of $X$~or $Y$~errors. Such schemes will likely need to be designed with the noise parameters of a specific device in mind. In particular, bias-sensitive error correction is critical for vulnerable sections of distillation schemes.

With that in mind, this work demonstrates reassuring results that bias can be high even when performing state injection in low-distance codes.
Although we used the Steane code, one example of a low-distance code, the noise was still $Z$-biased at the logical level.
We conjecture that the scaling of logical bias with distance found in this work will generalise to larger distance CSS codes, a potential area for research, but ultimately work in error correction will be necessary to allow for high-bias to manifest in magic state injection.
Further work that explores error correction with close consideration for its noise characteristics would be a promising area for future research.

\begin{acknowledgments}
We thank Evan Hockings for helpful comments on an early draft. This work is supported by the Australian Research Council via the Centre of Excellence in Engineered Quantum Systems (EQUS) project number CE170100009, and by the ARO under Grant Number: W911NF-21-1-0007. Access to high-performance computing resources was provided by the National Computational Infrastructure (NCI), which is supported by the Australian Government, and by the Sydney Informatics Hub, a Core Research Facility of the University of Sydney.
\end{acknowledgments}

\bibliographystyle{quantum}

\providecommand{\noopsort}[1]{}\providecommand{\singleletter}[1]{#1}%

\newpage

\onecolumn
\appendix

\section{Exploring Pauli Bias}\label{sec:biasdef}

\begin{figure*}[t]
\centering
\begin{flushleft}
\large (a)\hspace{0.5\textwidth}(b)\\
\end{flushleft}
\vspace*{-5mm}
$\vcenter{\hbox{\includegraphics[width=0.2\textwidth]{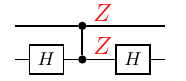}}} =\vcenter{\hbox{\includegraphics[width=0.105\textwidth]{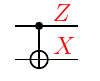}}}$
        \hfill    
$\vcenter{\hbox{\includegraphics[width=0.35\textwidth]{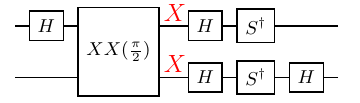}}} =\vcenter{\hbox{\includegraphics[width=0.105\textwidth]{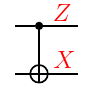}}}$
\caption{In~(a) $Z$ noise on a $C_Z$ becomes mixed noise $\mathcal{M} = \{Z_1,X_2,Z_1X_2\}$ when used to implement a CNOT. The final $\mathcal{M}$-bias $\eta_2^\mathcal{M}$ of the noise will equal the initial $Z$-bias~$\eta_2^Z$ of the $C_Z$ noise. The same applies for $X$-bias~$\eta_2^X$ in noisy MS-gates~\cite{MS}, as depicted in~(b).}
\label{fig:zxnoise}
\end{figure*}

We have used two-qubit bias throughout this work to parameterise physical noise. Different kinds of bias have been introduced to help identify which aspects of physical noise drive the bias of logical noise. Despite the seemingly ad-hoc manner that each bias has been defined, all biases mentioned have maintained consistent properties that we now explicate.

For all sets of high-rate errors used in this text~$\mathcal{Q}\subset\mathcal{P}_n\backslash\{\mathbb{I}\}$, any Pauli error $Q\in\mathcal{Q}$ commutes with all other Pauli errors in~$\mathcal{Q}$. The set is always chosen to have the largest possible cardinality, that is, $|\mathcal{Q}| = 2^n-1$ for $n$-qubit noise. We can express such high-rate sets in terms of $n$ independent commuting generators 
\begin{equation}\mathcal{Q} = {\langle Q^{(1)}, Q^{(2)}, Q^{(3)}, \dots, Q^{(n)} \rangle \backslash \{\mathbb{I}\}} \end{equation}
The corresponding set of low-rate errors $\mathcal{Q}^C$ has cardinality $|\mathcal{Q}^C| = 4^n-2^n$, such that $n$-qubit depolarising noise has a bias value of $\eta_n^\mathcal{Q}=|\mathcal{Q}|/|\mathcal{Q}^C|=2^{-n}$ for these choices of~$\mathcal{Q}$ because each Pauli error has equal probability.

In the same way that we extend single-qubit bias to two-qubit bias in Sec.~\ref{sec:bias}, we can straightforwardly generalise single-qubit bias to $n$-qubit bias
\begin{equation} \eta_n^P \vcentcolon= \eta_n^{\langle P_1, P_2, P_3, \dots, P_n \rangle \backslash \{\mathbb{I}\}} \end{equation}

$Y$-biased noise is distinct from $X$- and $Z$-biased noise in the context of faulty CNOTs. For the latter two, whether the noise occurs before or after the CNOT has no effect on the extent of bias or the rates of error for the Pauli noise models we considered. The CNOT permutes errors within~$\mathcal{Q}$ and~$\mathcal{Q^C}$ but does not exchange errors between the two sets. For $Y$-biased noise, errors in~$\mathcal{Q}$ can become errors in~$\mathcal{Q^C}$ when the Pauli noise channel occurs before the CNOT. Although $Y$-biased noise differs in this regard, simulations that considered $\eta_2^Y$ gave comparable results to $\eta_2^X$ and provided no substantial additional insight into the characteristics of logical noise.

The mixed-bias $\mathcal{M} = \{Z_1,X_2,Z_1X_2\}$ is a notable outlier among the two-qubit biases presented in this text, and there are various alternative candidates that could have been chosen for mixed bias. We were motivated by the following unique features of~$\mathcal{M}$: just as with $\eta_2^X$ and $\eta_2^Z$, the extent of $\mathcal{M}$-biased noise is unchanged if the biased noise channel occurs before or after the CNOT. In addition, $\mathcal{M}$~is the only high-rate set where all constituent errors commute with the CNOT operator, making it a fairly natural choice for mixed-bias in the context of faulty CNOTs. As Fig.~\ref{fig:zxnoise} suggests, $\mathcal{M}$-biased physical noise is related to other sensible kinds of bias motivated by device noise~\cite{mckay19, cong22, saha23, berm19}.

As a final note, the extent of bias $\eta_n^\mathcal{Q}$ is not fundamentally changed under the application of a Clifford gate on the same $n$-qubit space. For any defined high-rate set $\mathcal{Q}$ and Clifford gate $C$ we can define
$\mathcal{Q}' = \{ CQC^\dag | \forall Q \in \mathcal{Q}\}$
such that the extent of bias $\eta_n^\mathcal{Q}$ before the gate $C$ is equal to $\eta_n^{\mathcal{Q}'}$ after the gate~$C$. The Clifford frame may change, but the extent of bias (the numerical value) remains unchanged.

\begin{figure}[t]
\begin{center}
$\vcenter{\hbox{\includegraphics[width=0.06\textwidth]{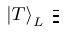}}} =\vcenter{\hbox{\includegraphics[width=0.27\textwidth]{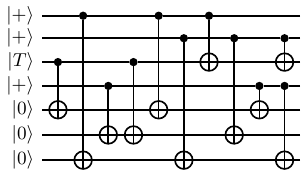}}}$
\end{center}
\caption{Encoding of the $\ket{T}$ state into the Steane code~\cite{cham19, goto16}. A local single-qubit fault on the unencoded $\ket{T}$ will be encoded as a logical error.}
\label{fig:t-state-prep}
\end{figure}

\section{State Preparation}\label{sec:state-prep}

We divide logical state preparation into three components: a non-fault-tolerant state preparation, a logical state check, and a round of stabiliser measurements performing error detection. Since stabiliser measurements apply more broadly than state preparation, we consider them separately in Sec.~\ref{sec:err-corr}. In this section we identify the critical errors from state preparation that the later error detection must address.

We will require the final logical state to have no more than a single-qubit error at first order because we aim to prepare the state fault-tolerantly and we are operating within the Steane code.
Without this requirement subsequent error correction may not be able to recover the ideal state at first order because the character of the Pauli errors can change throughout the circuit. We will keep this important constraint in mind when considering sources of noise throughout this section.

\begin{figure}[b]
\begin{center}
$\vcenter{\hbox{\includegraphics[width=0.06\textwidth]{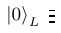}}} =\vcenter{\hbox{\includegraphics[width=0.2\textwidth]{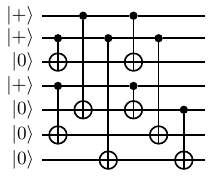}}}$
\end{center}
    \caption{Encoding of the $\ket{0}$ state into the Steane Code~\cite{goto16}. Without performing any additional error detection or correction, the prepared logical state may have errors of weight-one or weight-two.}
    \label{fig:state-prep}
\end{figure}

The schemes discussed in this section come from the work of Goto~\cite{goto16} and Chamberland \& Cross~\cite{cham19}. Here we consider how the structure of these schemes matter within the context of our work.

\subsection{Encoding Circuits}\label{sec:enc-circ}

Encoding circuits are a type of Clifford operator that take a number of unentangled qubits into the logical space of some desired code. These circuits are not unique and we do not attempt to improve upon existing encoding schemes here. We use encoding circuits as found in existing work~\cite{goto16, cham19}, which we depict in Figs.~\ref{fig:t-state-prep} and~\ref{fig:state-prep}. 

Simple encoding circuits are generally not fault-tolerant because the initial states are unencoded and thus vulnerable to errors. 
Physical errors on the initial states may get encoded into logical errors, and at the very least, local errors often spread and become higher weight errors.
One such case occurs when we prepare magic states using the encoding circuit of Fig.~\ref{fig:t-state-prep}, where errors on the third qubit can become logical errors. Logical errors cannot be corrected or detected by code stabiliser checks, demonstrating a need for other checks in order to prepare logical states fault-tolerantly.

This does not apply for all circuits. Some circuits that we use, such as the specific preparation of logical stabiliser states depicted in Fig.~\ref{fig:state-prep}, do not introduce undetectable logical errors from any single-qubit sources.
Local errors may spread into high-weight errors, but they remain detectable. Although not as necessary when compared to magic state preparation, additional logical checks are still useful for mitigating the problem of high-weight errors, especially when faced with correlated noise.

\begin{figure}[b!]
\begin{center}
\quad\quad\quad\quad
$\vcenter{\hbox{\includegraphics[width=0.06\textwidth]{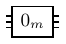}}} =\vcenter{\hbox{\includegraphics[width=0.16\textwidth]{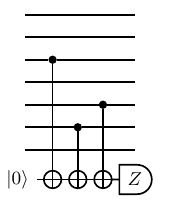}}}$
\hfill
$\vcenter{\hbox{\includegraphics[width=0.06\textwidth]{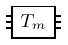}}} =\vcenter{\hbox{\includegraphics[width=0.315\textwidth]{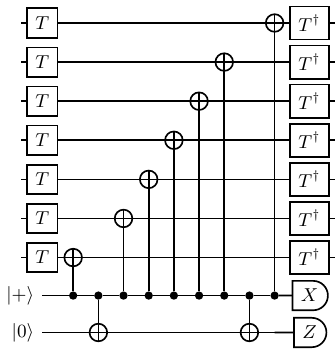}}}$
\quad\quad\quad\quad
\end{center}
    \caption{Logical check circuits that detect errors on the prepared logical state. These circuits are post-selected on +1, projecting the logical qubit onto the corresponding logical subspace. $0_m$ projects the state onto the +1 eigenspace of $Z_L$, while $T_m$ projects the state onto the +1 eigenspace of $(X_L+Y_L)/\sqrt{2}$, an operator that stabilises the logical magic state $\ket{T}_L$.}
    \label{fig:log-check}
\end{figure}

\subsection{Logical Checks}\label{sec:log-check}

Logical check measurements are a requirement for fault-tolerant state preparation in addition to the usual stabiliser checks. The previous subsection App.~\ref{sec:enc-circ} indicated why logical checks are necessary, although the specific problems that the checks need to address depend on the state being prepared. The checks we review here originate from existing literature~\cite{goto16, cham19} and are depicted in Fig.~\ref{fig:log-check}. In this section, we highlight the differences between these state preparation schemes and detail the main types of noise that they fail to detect.

The logical check used in the preparation of the $\ket{0}_L$ state must account for high weight $X$ noise. Due to properties of the $\ket{0}_L$ state, $Z$ noise is not a major concern: logical $Z$ errors act trivially on the state, also meaning that a weight-two $Z$ error is equivalent to a weight-one $Z$ error. Hence, Goto's check circuit does not attempt to detect $Z$ errors, detecting only $X$ errors instead. It does so with just one additional ancilla post-selected on $+1$. 

\begin{figure}[t]
\begin{center}
\includegraphics[width=0.32\textwidth]{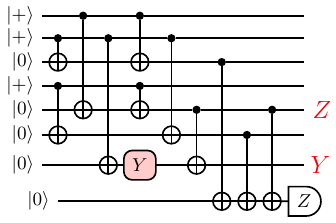}
\end{center}
    \caption{A fault-tolerant state preparation for logical $\ket{0}$ in the Steane Code \cite{goto16}. Post-selecting the ancilla measurement on $+1$ ensures that no single fault results in an undetectable logical error. Here a local error $Y_7$ spreads to $Z_5Y_7$. Note that $Z_L = Z_2Z_5Z_7$ and $Z_5Y_7 = Z_2 X_7 Z_L \sim Z_2 X_7$. By applying a suitable $Z_L$, all emerging weight-two errors may be expressed as either single-qubit $X$ and $Z$, or single-qubit $Y$ and $Z$.}
    \label{fig:state-prep-final}
\end{figure}

\begin{figure*}[t]
\centering
$\vcenter{\hbox{\includegraphics[width=0.07\textwidth]{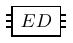}}} =\vcenter{\hbox{\includegraphics[width=0.83\textwidth]{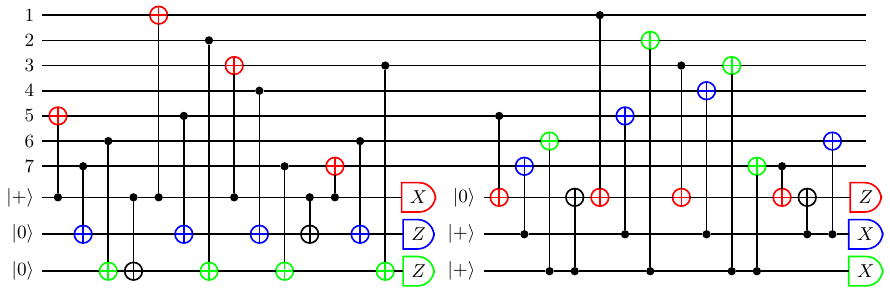}}}$

    \caption{A flagged scheme for performing error detection fault-tolerantly on the Steane code. The CNOTs for each stabiliser measurement are coloured red, green and blue corresponding to the stabiliser generators depicted in Fig.~\ref{fig:steane}. This scheme is a modification of the one introduced by Reichardt~\cite{reich18no} such that no weight-two errors over the support of any CNOT will amount to errors of weight-two or higher on the data qubits without also flipping an ancilla measurement. This stabiliser measurement scheme is used within this work for error detection during state preparation, however, the original scheme by Reichardt is used for stabiliser measurements performing error correction $EC$.}
\label{fig:fullstab}
\end{figure*}

For the Pauli noise models considered within this work, this check ensures that the state is prepared with no logical errors at first order; there can be no more than a single $Z$ error, and the logical check ensures that there is no more than a single $X$ error. However, this does not ensure that these errors are localised to a single qubit. This is particularly true when considering two-qubit errors introduced by CNOTs, however, as Fig.~\ref{fig:state-prep-final} shows, even errors that are initially localised to a single-qubit can become undetected multi-qubit errors.
Such weight-two errors are a problem when performing error correction because the Steane code's distance is only three. Given this, we will need to account for such errors with further stabiliser checks.
Although weight-two errors comprised of $Z$ and $X$ are correctable independently, such errors can change character as they propagate throughout the circuit and so require additional care to address as their syndromes change.
For our simulations we follow the state preparation with a round of error detection, outlined in Fig.~\ref{fig:fullstab}, which has been designed not to introduce any weight-two errors at first order.

For the magic state logical check, first note that the check circuit in Fig.~\ref{fig:log-check} is the same as the one provided by Chamberland \& Cross~\cite{cham19} but performed in a different basis. In their paper, they prepare the Hadamard eigenstate $\ket{H} = \cos(\pi/8)\ket{0}+\sin(\pi/8)\ket{1}$ which is injected to implement $Y(\pi/4)$. The $\ket{T}$ magic state is equivalent to the Hadamard eigenstate up to Clifford operations $\ket{H}= SH\ket{T}$, likewise, the check circuit $T_m$ is equivalent to their check circuit up to Clifford operations.

Unlike the preparation of $\ket{0}_L$, the preparation of $\ket{T}_L$ must deal with a more fundamental problem: logical errors occur at first order in the circuit of Fig.~\ref{fig:t-state-prep}. By measuring $(X_L+Y_L)/\sqrt{2}$ with the check circuit $T_m$ we can detect those logical errors and avoid them by post-selecting on $+1$. However, high-weight errors will still occur on the prepared state, some of which cannot be corrected. Weight-two $Z$ errors commute with the check and thus cannot be detected by it.
Weight-two $X$ and $Y$ errors will remain as weight-two errors after post-selecting on $+1$.
There are also faults during $T_m$ that propagate to the data qubits.
Therefore, we must follow the magic state preparation scheme with a round of error detection.

Both state preparation schemes rely on a single successful logical check to be fault-tolerant. Therefore, whenever the check's ancilla has a measurement error, other errors can go entirely undetected. 
This means that the state preparation schemes used here perform no better than second order. 

It is for that reason that we do not consider logical tomography of a logical $T$-gate where errors are being error detected.
While the lowest order logical errors from the transversal gates will be third order due to error detection, the logical errors of state preparation will be second order.
Therefore, the bulk of the logical noise would come from state preparation and overshadow the actual noise from state injection.
The results of such a tomography would reflect the noise associated with state preparation rather than the pertinent characteristics of the entire injection scheme.

\section{Error Correction}\label{sec:err-corr}

Error correction is a crucial component to all fault-tolerant schemes. 
Without error correction the ongoing faults of a device will accumulate, making any quantum information unusable.
Transversal gates help to ensure that errors on a device do not spread as quickly; errors that are correctable before a transversal gate should remain correctable after a transversal gate.
However, these transversal gates must still be followed by a round of stabiliser measurements in order to take advantage of the correctability of the errors. 
As seen in the previous section, these stabiliser measurements are also necessary to prepare logical states fault-tolerantly. 

Here we precisely detail how error correction is implemented within our simulations, with specific focus on the complications introduced by correlated noise. This close analysis is particularly motivated by error correction's pivotal role in constraining the extent of logical bias.

Although errors composed of a single $Z$ error and a single $X$ error can be independently corrected in the Steane code, the appropriate correction for these errors may be dependent on the logical gates that follow.
Under the application of a transversal gate, such as the $S$-gate, these errors can go from~$ZX$~to~$ZY$ which is uncorrectable unless we alter the correction to take into account this change of character. 
Hence we must carefully account for such faults since we cannot unequivocally correct them without also accounting for how they will propagate through logical gates.
If there are independent sources of such correlated errors, but only some of them are converted to $ZY$ while others remain as $ZX$, then we will be unable to adopt a correction strategy that can accommodate both sources, meaning that one of these errors will become a logical error after applying the correction.
Therefore, when taking for granted that particular circuit elements have correlated $ZX$ faults, we need to be careful that all such errors remain correctable in the context of logical gates.
We take particular efforts to avoid such faults as much as possible.

\begin{figure*}[t]
\centering
$\vcenter{\hbox{\includegraphics[width=0.07\textwidth]{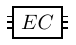}}} =\vcenter{\hbox{\includegraphics[width=0.83\textwidth]{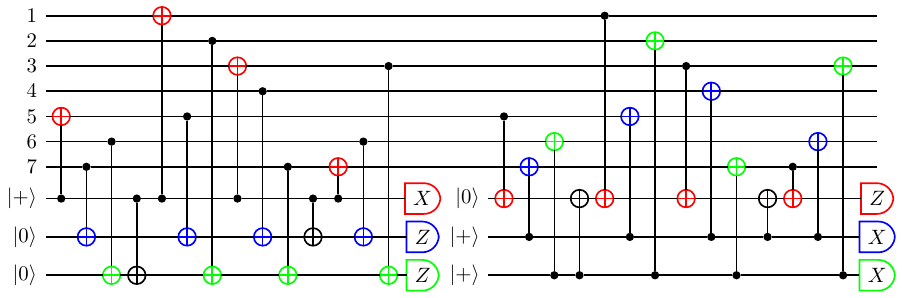}}}$

    \caption{A flagged scheme for performing error correction fault-tolerantly on the Steane code~\cite{reich18no}. The CNOTs for each stabiliser measurement are coloured red, green and blue corresponding to the stabiliser generators depicted in Fig.~\ref{fig:steane}. A modified version of this scheme, used for error detection, is presented in Fig.~\ref{fig:fullstab}.}
\label{fig:EC}
\end{figure*}

\subsection{Correlated Hook Errors}

When preparing logical states in the Steane code, we must ensure that any first order errors that escape detection are localised to a single-qubit. If that is not the case, the errors may later get corrected into logical errors because the Steane code's distance is only three. This is a substantial challenge when we consider correlated noise, since the weight-two errors occurring in non-fault-tolerant components must not affect any more than a single qubit of the final state.

Earlier work~\cite{cham19} has utilised Reichardt's flagged measurement scheme~\cite{reich18no} to prepare logical states in the Steane code. Using no additional ancilla qubits, this flagged scheme is able to deal with a type of fault called a \textit{hook error}~\cite{yoder17, reich18no, reich18two}. The flag CNOTs of Reichardt's scheme ensure that hook errors can be detected and corrected appropriately. Without using flags, single-qubit hook errors are able to spread to multiple data qubits undetected. 

\begin{figure}[t]
\begin{center}
\includegraphics[width=0.32\textwidth]{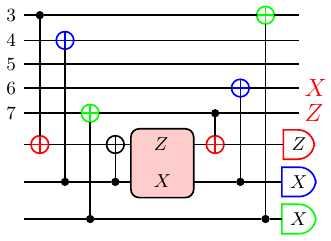}
\end{center}
\caption{Correlated errors in flagged schemes can spread to weight-two errors on the logical qubit. Depicted here is a section of a flagged stabiliser measurement scheme by Reichardt~\cite{reich18no}. A correlated error introduced to the ancilla qubits by the flag CNOT spreads undetected to the computational qubits as a weight-two error $X_6Z_7$.}
\label{fig:reichfail}
\end{figure}

However as Fig.~\ref{fig:reichfail} demonstrates, there are certain errors that this scheme fails to tolerate when considering correlated weight-two errors over a single CNOT.
Such errors can occur without flipping an ancilla measurement, allowing weight-two $ZX$ errors to spread to the logical state undetected by the flag circuit, which we want to avoid if possible.
When using this stabiliser measurement for state preparation, such sources of correlated noise compromise the fault-tolerant preparation.

This kind of multi-qubit hook error is a problem specific to flagged schemes. By applying CNOTs between ancilla qubits, we introduce a new way that correlated noise may arise. High-weight errors may now occur correlated between multiple ancilla qubits, rather than independently, leading to a new kind of multi-qubit hook error that can be a problem for fault-tolerance.

In order to account for such errors, we modify this flagged scheme for use in state preparation. Our modified circuit is depicted in Fig.~\ref{fig:fullstab}. Rearranging only a few CNOTs, we can detect all offending correlated errors from within the stabiliser measurement such that only single-qubit errors evade error detection. 

By using the circuit in Fig.~\ref{fig:fullstab} in conjunction with the state preparation components of App.~\ref{sec:state-prep}, we can prepare logical states with at most single-qubit errors at first order. We use this specific setup in all our simulations.

\subsection{Correlated Errors and Leakage}\label{sec:leak}

Correlated errors are particularly troublesome in low-distance codes because correlated errors may have a weight similar to the code distance, so any indirect mechanisms that introduce errors that are correlated across the data qubits can not generally be tolerated.
Such issues drive us to integrate error correction into logical tomography itself, which we will discuss here.

\begin{figure}[b!]
\begin{center}
\includegraphics[width=0.48\textwidth]{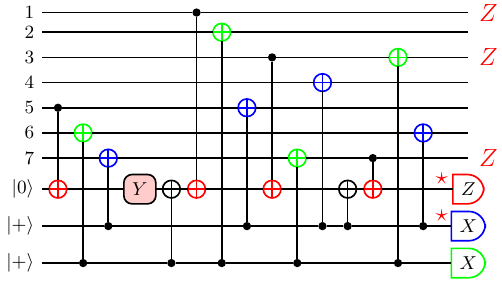}
\end{center}
\caption{A single-qubit error can become a high-weight error after performing error correction. Although it appears as if the $Y$~error has already spread into a high-weight error, $Z_1Z_3Z_7 = Z_5 Z_R$, where $Z_R$ is the red $Z$~stabiliser. So effectively only a single error $Z_5$ has been introduced. Note that the $Z$~error introduced to qubit~3 also spreads to the green ancilla via the final green CNOT such that the $Y$~error triggers two ancilla measurements instead of three. After applying the corresponding correction the final error is $X_1Z_4Z_5 = Y_1Z_L$. Such an error goes undetected by logical $Y$ measurements but acts as a logical error. $Y_1Z_L$ is still detectable when performing $X$ and $Z$ syndrome measurements.}
\label{fig:stab-singleerror}
\end{figure}

There are various mechanisms by which a weight-two error arising during stabiliser measurements can result in a weight-two error on the data qubits.
Setting aside flag CNOTs, correlated noise on the regular CNOTs of a stabiliser measurement will introduce one error to a data qubit and one error to an ancilla qubit. The ancilla error may be detected, prompting a subsequent correction corresponding to the syndrome.
For many such cases the correction will be applied to a different qubit than the pre-existing data error.
Although the initial weight-two error does not affect two data qubits, it can after error correction. For our noise models, this would mean that weight-two errors are introduced by faulty error correction at first order.

Even without high-weight errors, single-qubit errors may cause problems for logical tomography, as outlined in Fig.~\ref{fig:stab-singleerror}.
The error in the figure is initially a local $Y$~error, spreading to the data qubits as a single $Z_5$~error while flipping two stabiliser checks.
After error correction it is equivalent to an error on three separate qubits, anticommuting with logical $Y$, but commuting with all $Y$ stabilisers. 
Hence this singular fault results in an undetectable logical measurement error after the relevant $ZX$ correction is applied.
Correlated weight-two errors in error correction cause similar problems for tomography.

This error is not a logical operator, therefore, we can still detect the error if we are careful with our measurements.
For that reason, we change the logical basis for the tomography measurement by applying $C^{\otimes7}$ before error correction, rather than after, so that such errors can be detected by the final transversal measurement and discarded as leakage from the logical space.

In general, a fault-tolerant implementation of magic state injection would follow up the stabiliser measurement with additional measurement rounds.
Here we instead use information gained from the final tomographic measurement to detect additional errors, but we note that this information gained in tomography is not part of the logical process being characterised.

\section{Logical Tomography}\label{sec:tomography}

In order to perform fault-tolerant logical tomography we must ensure that logical state preparation and logical measurement are implemented fault-tolerantly, otherwise logical SPAM errors will dominate the noise of the derived logical process. We also need to perform error correction to ensure the process itself is fault-tolerant. We discuss and define these specific circuit components in other parts of the appendix: fault-tolerant state preparation in App.~\ref{sec:state-prep} and stabiliser measurements in App.~\ref{sec:err-corr}. Here we focus on how the measurement outcomes of our simulations are used to perform logical tomography.

\begin{figure}[b]
    \begin{center}
    \includegraphics[width=0.48\textwidth]{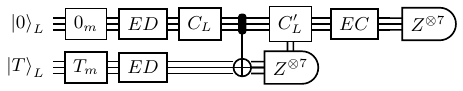}
    \end{center}
    \caption{Logical circuit structure for our simulations of logical $T$-gates implemented through magic state injection. The circuit is encoded in the Steane code. We modify the input stabiliser state and final measurement basis of the circuit by varying the logical Clifford operators $C_L$ and $C'_L$, allowing us to perform logical process tomography. The final measurements are transversal applications of physical $Z$ measurements whose parity is used to determine the logical measurement outcome.}
    \label{fig:full-circ}
\end{figure}

The logical circuits that we simulate are depicted in Fig.~\ref{fig:full-circ}, from which we derive the full assortment of circuits required to perform tomography by varying the Clifford operators $C_L$ and $C'_L$. To prepare initial states $\ket{0}_L$, $\ket{1}_L$, $\ket{+}_L$ and $\ket{+_Y}_L$, $C_L$ is chosen to be $I^{\otimes7}$, $X^{\otimes7}$, $H^{\otimes7}$ and $(S^\dag H)^{\otimes7}$ respectively. We also need to measure in the three Pauli bases, $X$, $Y$ and $Z$, however we incorporate the adaptive $S_L$ correction into $C'_L$ to keep in mind circumstances where adaptive control is not possible. In those circumstances six circuits are required per initial state rather than three. For $X$, $Y$ and $Z$ logical measurements with logical magic state outcome of 0, $C'_L$ is chosen to be $H^{\otimes7}$, $(HS)^{\otimes7}$ and $I^{\otimes7}$ respectively. For $X$, $Y$ and~$Z$ logical measurements with logical magic state outcome of 1, $C'_L$ is chosen to be $(HS^\dag)^{\otimes7}$, $H^{\otimes7}$ and $(S^{\dag})^{\otimes7}$ respectively. Overall this amounts to 12 distinct circuits if adaptive circuits can be performed, increasing to 24 if they cannot.

As discussed in App.~\ref{sec:state-prep} we initially use post-selection to ensure the fault-tolerant preparation of logical states.
Subsequently we use stabiliser measurements to perform error correction. 
We also use the transversal $Z$~measurement of the logical magic state to correct $X$~errors, using the products of measurement outcomes to simultaneously infer both the logical $Z$ outcome and the $Z$~stabiliser outcomes. 
This ensures that the logical process is deterministic. 

The final transversal $Z$ measurement of the computational qubit is not handled in the same way since this measurement is expressly being used for logical tomography, not for error correction. 
We need the logical measurement outcome to be tolerant to any single fault from the circuit.
However, there are single faults that can occur within the stabiliser checks that, after applying error correction, cause the logical measurement to fail due to the low distance of the Steane code.
These errors are still detectable by the final measurement, which should otherwise report a trivial syndrome because the state should have been returned to the logical space.
Hence, after performing error correction any further errors detected are due to stabiliser faults, and are treated as leakage from the logical space.
App.~\ref{sec:leak} discusses these leakage errors in more detail.
Note that although this is a suitable way to perform fault-tolerant logical measurements for tomography in the Steane code, it should not be used for codes with a large number of qubits. Nonetheless, larger codes should also have a large distance, and so would be able to tolerate these kinds of faults anyway.

The logical probabilities obtained via simulation of each distinct circuit are then used to derive the logical process matrix by using the generalised inverse~\cite{niel11}, also commonly referred to as the pseudoinverse.

\end{document}